\documentclass[11pt]{amsart}

\newtheorem{thm}{Theorem}[section]
 \newtheorem{cor}[thm]{Corollary}
 \newtheorem{lem}[thm]{Lemma}
 \newtheorem{prop}[thm]{Proposition}
 \theoremstyle{definition}
 
 \newtheorem{examp}[thm]{Example}
 \theoremstyle{remark}
 \newtheorem{rem}[thm]{Remark}
 \theoremstyle{notation}
 \newtheorem{nota}[thm]{Notation}
 \numberwithin{equation}{section}

  \newcommand{\s}{S}
 
 \newcommand{\f}{\mathcal{F}}
  \newcommand{\Z}{\mathbb{Z}}

 \newcommand{\cc}{\mathcal{C}}
 
  \newcommand{\h}{\mathfrak{h}}
 \newcommand{\bneg}{\mathfrak{b}^-}
  \newcommand{\bpos}{\mathfrak{b}^+}

\newcommand{\ngauge}{N^-}
\newcommand{\nneg}{\mathfrak{n}^-}
\newcommand{\gneg}{\mathfrak{g}^-}
\newcommand{\gpos}{\mathfrak{n}^+}
 \newcommand{\g}{\mathfrak{g}}
 \newcommand{\Ad}{\textrm{Ad}}
 \newcommand{\ad}{\textrm{ad}}
 \newcommand{\e}[1]{\textrm{exp}(\textrm{ad}~#1)~}
 \newcommand{\lcan}{L^{\!c}}
 
\newcommand{\gauge}{G^-}

 \newcommand{\lop}[1]{\mathfrak{L}(#1)}
 \newcommand{\w}{\mathbf{W}}
 \newcommand{\orb}{\mathcal{O}}
\begin{document}

\title[On classification of algebraic Frobenius manifolds]
 { On classification and construction of algebraic Frobenius manifolds
 }

\author{Yassir Ibrahim Dinar }

\address { University of Khartoum, Khartoum, Sudan
}
\email{dinar@ictp.it}


\begin{abstract}
We develop the theory of generalized
bi-Hamiltonian reduction. Applying this theory to a suitable loop algebra
we recover a generalized Drinfeld-Sokolov
reduction. This gives a way to construct new examples of algebraic
Frobenius manifolds.
\end{abstract}
\maketitle
\tableofcontents
\section{Introduction}
This work was intended as an attempt to prove the
Dubrovin conjecture \cite{DuRev} (see also \cite{DMD}).\\
\textbf{The conjecture}: Massive irreducible algebraic Frobenius manifolds
with positive degrees $d_i$ correspond to primitive conjugacy
classes in Coxeter groups.\\
A \textbf{Frobenius manifold} is a manifold $M$ with
the structure of Frobenius algebra on the tangent space $T_t$ at
any point $t \in M $ with certain compatibility conditions \cite{DuRev}. We
say $M$ is \textbf{massive} if $T_t$ is semisimple for generic
$t$. This structure  locally correspond to a potential
$\mathbb{F}(t^1,...,t^n)$ satisfying the WDVV equations
\begin{equation} \label{frob}
 \partial_i
\partial_j
\partial_k \mathbb{F}(t)~ \eta^{kp} ~\partial_p
\partial_q
\partial_r \mathbb{F}(t) = \partial_r
\partial_j
\partial_k \mathbb{F}(t) ~\eta^{kp}~\partial_p
\partial_q
\partial_i \mathbb{F}(t)
  \end{equation}
where $(\eta^{-1})_{ij} = \partial_n \partial_i \partial_j
\mathbb{F}(t)$ is a constant matrix. Here we assume that the quasihomogeneity condition takes the form
\begin{equation}
\sum_{i=1}^n d_i t_i \partial_i \mathbb{F}(t) = \left(3-d \right) \mathbb{F}(t)
\end{equation}
where $d_n=1$. This condition defines the degrees $d_i$ and the charge $d$ of   $M$. If $\mathbb{F}(t)$ is an algebraic function we call $M$ an
\textbf{algebraic Frobenius manifold}.

 A Coxeter group is a finite group of linear transformations acting on an Euclidean space generated by reflections \cite{HumCOX}. Irreducible Coxeter groups are classified by a set of reflections, called \textbf{simple reflections}, that generate all the group. The Weyl group of a simple Lie algebra is an irreducible Coxeter group. A \textbf{primitive conjugacy class } in a Coxeter group  is a  conjugacy class such that writing any representative of the class as a product of reflections, these reflections generate all the group \cite{CarClassif}. The set of simple reflections  of a Coxeter group defines  a conjugacy class called \textbf{Coxeter conjugacy class}. The conjugacy class is called \textbf{regular} if it has a regular eigenvector, i.e. an eigenvector which is not fixed by any element of the group \cite{springer}.

The Dubrovin conjecture arises from algebraic solutions to equations of isomonodromic deformation of algebraic Frobenius manifolds \cite{DuRev}.  The classification of  finite orbits of the braid group action on tuple of reflections obtained by Stefanov \cite{STF} (see also \cite{MIC}) means that algebraic Frobenius manifold leads to a primitive conjugacy class in  Coxeter groups. Thus it remains the problem of constructing  an algebraic  Frobenius manifold for any primitive conjugacy class in Coxeter groups.

Hertling  \cite{HER} proved that any irreducible massive \textbf{polynomial Frobenius manifold} with
positive degrees $d_i$ is isomorphic to the Frobenius structure
defined by Dubrovin on the
orbit spaces of a Coxeter group \cite{DCG}. The polynomial Frobenius manifold corresponds to  Coxeter conjugacy class in the group \cite{DuRev}.

Our main idea is to use the theory of
infinite dimensional bi-Hamiltonian manifolds to construct all
massive algebraic Frobenius manifolds. \textbf{A bi-Hamiltonian manifold} is a manifold endowed with two Poisson tensors $P_1$ and $P_2$ such that
$P_\lambda=P_2+\lambda P_1$ is a Poisson tensor for any constant
$\lambda$. The dispersionless limit of a bi-Hamiltonian structure
on the loop space $\lop M$ of a finite dimensional manifold $M$ (if it  exists) always
gives a bi-Hamiltonian structure of hydrodynamic type:
\begin{equation}\{t^i(x),t^j(y)\}_{1,2} = g^{ij}_{1,2}(t(x))
  \delta'(x-y) + \Gamma^{ij}_{1,2;k}(t(x)) t^k_x \delta(x-y),\end{equation}
defined on the loop space $\lop M$. This in turn gives a flat pencil of
metrics $g^{ij}_{1,2}$ on $M$ which under some assumptions
corresponds to a Frobenius structure on $M$  \cite{DFP}.

We use the theory of Lie algebras to associate a bi-Hamiltonian structure to a conjugacy class in  Weyl groups.
Let $\g$ be a simple Lie algebra. Denote by $\w_\g$ the  Weyl group of $\g$. Let $[w]\subset \w$ be a primitive regular conjugacy class in $\w_\g$. From \cite{DelFeher} there exists a  nilpotent orbit $\orb_e\subset\g$ corresponding to $[w]$ in the sense of \cite{springer}. We define a compatible Lie-Poisson brackets on the loop space $\lop \g$ depending on a nilpotent element $e \in\orb_e$. Fix a transversal subspace $M$ to $\orb_e$ at the point $e$. Then we construct a bi-Hamiltonian structure on $\lop M$ by using a generalized bi-Hamiltonian reduction.

The bi-Hamiltonian structure on $\lop M$ does not always admit a dispersionless limit. In this case we perform a Dirac reduction \cite{MRbook} on a suitable submanifold $N\subset M$ to obtain a new bi-Hamiltonian structure on the loop space $\lop N$. The bi-Hamiltonian structure on $\lop N$ admits a dispersionless limit.

A generalized Drinfeld-Sokolov reduction is another procedure to obtain a bi-Hamiltonian structure on $\lop M$. We prove the two reductions are equivalent in the sense that both of them satisfy the hypotheses of Marsden-Ratiu theorem and give the same reduced structures. The equivalence was obtained by Pedroni \cite{Pedroni2} in the special case of the reduction associated with the principal nilpotent
element.

The classical $W$-algebras and their primary fields were  obtained in \cite{BalFeh} by using the generalized Drinfeld-Sokolov reduction. We obtain this result by applying the generalized bi-Hamiltonian reduction. In \cite{CASFAL} they construct the classical $W_n$-algebra of the Lie algebra of type $A_n$ by studying the relation between  bi-Hamiltonian and Drinfeld-Sokolov reductions \cite{Pedroni2}. Our method is more straightforward and does not depend on a particular type of the Lie algebra.

The paper is divided into four parts. In the next section we develop the theory of generalized bi-Hamiltonian reduction. The idea goes back to \cite{CMP} where a bi-Hamiltonian reduction is given  for every bi-Hamiltonian manifold, using Marsden-Ratiu theorem, by taking
 a level surface $\s$ of all the Casimirs of the first Poisson brackets and a
  distribution $D$ defined by the second Poisson bracket.

Section \ref{notation} contains a brief summary of the theory of nilpotent elements and gradings on a simple Lie algebra and we set up notations and terminology. In section 3.2 we apply the generalized bi-Hamiltonian
reduction to Lie-Poisson bracket on a loop algebra of simple Lie
algebra. We give a general bi-Hamiltonian reduction for any
nilpotent element $e$ with an associated
good grading (in the sense of \cite{ELC}).  In section 3.3 we indicate how the bi-Hamiltonian reduction associated with a nilpotent element may be used to obtain the primary fields of classical $W$-algebras.

Section 4 provides a detailed exposition of a generalized Drinfeld-Sokolov reduction which is a special case of the more general Drinfeld-Sokolov reduction scheme given in \cite{BalFeh1}. In section 4.1 we establish  the equivalence between the generalized bi-Hamiltonian  and generalized Drinfeld-Sokolov reductions.

Finally section 5 is devoted to our main aim which is constructing new examples of algebraic Frobenius manifolds. We write in section 5.1 the formulas for  Dirac reduction of infinite dimensional Poisson bracket on a loop space $\lop M$ to $\lop N$ where $N$ is a suitable submanifold of $M$. In  section 5.2 we apply bi-Hamiltonian reduction to the distinguished nilpotent elements in the Lie algebra of type $F_4$.  The reduced bi-Hamiltonian structures  give  (after Dirac reduction) four algebraic Frobenius  manifolds in agreement with Dubrovin's conjecture in the sense that the degrees and charge of the algebraic Frobenius manifold can be read from the eigenvalues and order of the corresponding regular conjugacy class.

\section{Bi-Hamiltonian Reduction and Transversal Manifolds}
We recall the Marsden-Ratiu reduction theorem for Poisson
 manifolds. For  more information one can consult \cite{MR}.
\begin{thm}\label{mar:rat} Let $M$ be a Poisson manifold with Poisson bracket
$\{.,.\}^M$. Let
 $S$ be a submanifold of $M$  with $i_s:S \to M$ the
 canonical immersion of $S$ in $M$. Assume $D$ is a distribution
 on $M$ satisfying:
 \begin{enumerate}
    \item $E=D\cap TS $ is an integrable distribution of
        $S$.

    \item The foliation induced by $E$ on $S$ is regular,
        so that $N=S/E$ is a manifold and $\pi :S
        \rightarrow N$ is a submersion.
    \item The Poisson bracket of functions of $F,G$ that
        are constant along $D$, is constant along $D$
    \item $P(D^0)\subset TS+D$, where $D^0$ is the
        annihilator of D.

 \end{enumerate}
 Then $N$ is a Poisson manifold with bracket $\{.,.\}^N$ given by
    \begin{equation}
    \{f,g\}^N\circ\pi=\{F,G\}^M\circ i_s,
\end{equation}
where F,G are extensions of $f\circ \pi$, $g\circ \pi$ that are
constant along $D$. \\
\end{thm}

 The following sections depend on this corollary which replaces
  the study of the quotient manifold $N$ with
the study of a submanifold in $S$.
\begin{cor}\label{mar:rat:cor} Replace the condition
$(2)$ by the condition\\
$(2)'$ There exist a transversal submanifold $Q$ to the
distribution $D$ on $S$, i.e. at any point $q\in Q$ we have
\[T_q S= E_q \oplus T_q Q\] Then $Q$ is a Poisson manifold with
 bracket $\{.,.\}^Q$ defined by \[\{f,g\}^Q=\{F,G\}^M \circ
i_Q\] where $i_Q: Q\hookrightarrow M$ is the canonical
immersion and $F,G$ are functions on $M$ extending $f,g$ and
constant along $D$.
\end{cor}

A bi-Hamiltonian manifold $M$ is a manifold endowed with two
Poisson tensors $P_1$ and $P_2$ such that $P_\lambda=P_2+\lambda
P_1$ is a Poisson tensor for any constant $\lambda$. The Jacobi
identity for $P_\lambda$ gives the relation
\begin{multline}\label{bih:cond}
    \{\{F,G\}_1,H\}_2+\{\{G,H\}_1,F\}_2+\{\{H,F\}_1,G\}_2+\\
    \{\{F,G\}_2,H\}_1+\{\{G,H\}_2,F\}_1+\{\{H,F\}_2,G\}_1=0
\end{multline}
for any functions $F,G$ and $H$ on $M$. The main implication of this identity is that the set of Casimirs of $P_1$ is a Lie algebra with respect to $P_2$. Our basic assumption is the following. There is a set
\begin{equation}
\Xi=\{K_1,K_2,...,K_n\}\end{equation} of independent Casimirs of $P_1$ ($n$ is not necessary equal to the corank of $P_1$) closed with respect to $P_2$. Let us denote by $\s$ a level set of $\Xi$ and
define the integrable distribution $D$ on $M$ generated by the
Hamiltonian vector fields \begin{equation}
X_{K_i}=P_2(dK_i),\qquad i=1,...,n.
\end{equation}
The following lemma says that $P_\lambda,~S$ and $D$ verify the hypotheses of Marsden-Ratiu theorem \ref{mar:rat},  except condition (2).
\begin{lem}\label{bi-Hamiltonian reduction} For any constant
$\lambda$
\begin{enumerate}
\item The functions which are constant along $D$ form a Lie
subalgebra with respect to $P_\lambda$. \item $v \in D^0$ if and
only if $P_\lambda(v) \in TS$. Here $D^0\subset T^*M$ is the
annihilator of $D$.
\end{enumerate}
\end{lem}
\begin{proof}The first condition is easily deduced from the relation
\eqref{bih:cond} and Jacobi identity for $P_2$. Since $P_1(T^*M)\subset TS$ the statement  $(2)$ is equivalent to proving that \[v \in D^0~~\textrm{if and
only if  }P_2(v) \in TS.\] To this end, let $v \in D^0$. Then
\begin{eqnarray}
  (v,D)=0 & \Longleftrightarrow & (v,P_2(dK_i))=0,~i=1,...,n \\
\nonumber   & \Longleftrightarrow & (P_2(v),dK_i)=0,~ i=1,...,n \\
 \nonumber  & \Longleftrightarrow & P_2(v)\in TS,
\end{eqnarray}
and the proof is complete.
\end{proof}
In the remainder of this section we assume there is a submanifold $Q \subset S$ transversal to $E=D
\cap TS$, i.e.
\begin{equation}  T_q S=E_q
\oplus T_q Q,\quad\textrm{for all}~q\in Q.
\end{equation}
 Following \cite{CMP}, $Q$ has a natural
bi-Hamiltonian structure $P_1^Q$, $P_2^Q$ from $P_1$, $P_2$
respectively (see also corollary \ref{mar:rat:cor}). Let $i:Q\hookrightarrow M$ be the
canonical immersion. Then the pencil $P_\lambda^Q$ is defined, for any functions
$f,g$ on $Q$, by
\begin{equation}
\{f,g\}_\lambda^Q=\{F,G\}\circ i
\end{equation} where
 $F, G$
are functions on $M$ extending $f,g$ and constant along $D$.

Our next purpose is to find a way to write the reduced Poisson pencil tensor. Here the advantage of having a transversal manifold $Q$ becomes clear.  \begin{lem}\label{Self:Consistency}
 For any  $q\in Q$ and $w \in T_q^*Q$ there exists $v\in T_q^* M$
 such that:
\begin{enumerate}
\item $v$ is an extension of $w$, i.e. $(v,\dot{q})=(w,\dot{q})$
for any $\dot{q}\in T_q Q$. \item $P_\lambda (v) \in T_q Q$, i.e.
$(v,P_\lambda(TQ)^0)=0$. \end{enumerate} Then the Poisson tensor
$P_\lambda^Q(w)$ is  given by\begin{equation} P_\lambda^Q
w=P_\lambda v \end{equation} for any extension $v$ satisfying
conditions (1) and (2).
\end{lem}
\begin{proof} $w \in T_q^* Q$ has an extension $v \in T_q M$
satisfying (2) if $T_q Q\cap P_\lambda(TQ)^0=0$.  Assume
$\dot{q}\in T_q Q \cap P_\lambda(TQ)^0$. Then
$\dot{q}=P_\lambda(r)$ for $r \in (TQ)^0$. Since $P_\lambda(r)\in
T_qS$, by lemma \ref{bi-Hamiltonian reduction} we have that $r \in D^0$. Then
\begin{equation} r\in (TQ)^0\cap D^0 \subset (TQ+D)^0 \subset
(TS)^0.
\end{equation}
This implies $P_\lambda(r)\in D$ which gives $\dot{q} \in E$. But
$\dot{q} \in T_q Q$. This shows that $\dot{q}=0$ and proves the first part.
Let $v_1, v_2\in T_q^*M$ be extensions of $w_1,w_2\in T_q^* Q$
satisfying the condition (2). Then
\begin{eqnarray}
  (w_1,P_\lambda ^Q w_2) &=& (v_1,P_\lambda v_2) \\\nonumber
  &=& (w_1,P_\lambda v_2)
\end{eqnarray}
where the first equality is obtained by definition and the second
one follows from condition (2).
\end{proof}

\section{Examples for Lie-Poisson brackets}

\subsection{Nilpotent elements and gradings in Lie algebras}\label{notation}
Here we introduce some notations and basic facts from the theory of nilpotent elements in  simple Lie algebras.

Let $\g$ be a simple Lie algebra over complex numbers
with a nondegenerate invariant bilinear form $\langle.|.\rangle$. For a vector subspace $V \subset \g$ we
denote by $V^\perp$ its orthogonal complement and by $\lop V$ its
loop space, i.e. the space of smooth maps
from the circle $S^1$ to $V$.

Introduce the following bilinear form on the loop algebra $\lop\g$:
\begin{equation} (u|v)=\int_{S^1}\langle u(x)|v(x)\rangle dx,~ u,v \in \lop M.\end{equation}
Then identify $(\lop\g)^*$ with $\lop\g$ using this bilinear form. For a functional $F$ on $\lop\g$ we
define the gradient $\delta H (q)$ to be the unique element in
$\lop\g$ such that
\begin{equation}
\frac{d}{d\theta}F(q+\theta
\dot{s})\mid_{\theta=0}=\int_{S^1}\langle\delta F|\dot{s}\rangle dx
~~~\textrm{for all } \dot{s}\in \lop\g.
\end{equation} We introduce the following Poisson tensors
\begin{eqnarray} \label{Lie:poisson}
P_2(v)&=& v_x+[q,v] \\
\nonumber P_1(v)&=& [a,v]
\end{eqnarray}
given at a point $q\in \lop \g$ and for every $v\in(\lop\g)^*$, here $a\in \g$ is constant element. It is well known that the pair in \eqref{Lie:poisson} defines a bi-Hamiltonian structure on $\lop\g$ \cite{MRbook}. Our first examples of bi-Hamiltonian reduction will be constructed from \eqref{Lie:poisson} by choosing an appropriate element $a\in \g$ and a set of Casimirs of $P_1$.

\label{integral gradation} Let
\begin{equation}\label{grad1}
\g=\oplus_{i\in \mathbb{Z}} \g_i
\end{equation}
be a $\mathbb{Z}$-grading on $\g$, i.e. $[\g_i,\g_j]\subset\g_{i+j}$ (We will omit the letter $\Z$ since any grading considered here is a $\Z$-grading). Since all derivations of $\g$ are inner this grading is defined as eigenspaces of $\ad~\widetilde{h}$ for some element $\widetilde{h}$.
\begin{equation}
\g_i=\{a\in \g|\ad~ \widetilde{h}(a)=i a\};\end{equation} Hence $\widetilde{h}$ is semisimple.

An element $e\in \g_2$ is called \emph{good} if it
satisfies the following condition;
\begin{equation}\label{injNil} \textrm{ad}~e :\g_j\to \g_{j+2} \textrm{ is
injective for }  j\leq -1.
\end{equation}
A grading is called \emph{good} if it admits a
good element. All good gradings on simple Lie algebras up to conjugation  are classified
in \cite{ELC}.
\begin{nota} For any good grading we introduce the following subalgebras;
$\bneg=\oplus_{i\leq 0} \g_i$, $\bpos= \oplus_{i\geq 0}\g_i$, $\nneg=\oplus_{i\leq
{-1}}\g_i$, $\gneg=\oplus_{i\leq -2} \g_i$ and
$\gpos=\oplus_{i\geq 1}
\g_i$. \end{nota}

A subspace $\cc \subset \g $ is called a \emph{transversal
subspace} of $e$ if
\begin{equation}\textrm{ad}e(\gneg) \oplus \cc=\bneg.
\end{equation}

Let $e \in \g$ be an arbitrary
nilpotent element. By Jacobson-Morozov theorem there exist $h$ and $f \in \g$
such that $\{e,h,f\}$ is an $sl_2$-triple, that is,
\begin{equation}\label{sl2:relation} [h,e]=2 e,\quad [h,f]=-2
f,\quad [e,f]= h.
\end{equation}
 From representation theory of $sl_2$ it is easy to see that $h$ defines a grading on $\g$ with $e$ a good element. We call the grading thus obtained the Dynkin grading. A nilpotent orbit is the conjugacy class of a nilpotent element under the action of the adjoint group. It turns out that two nilpotent elements are conjugate if and only if they have the same Dynkin grading.  See \cite{COLMC} for more information and the classification tables of the nilpotent orbits which are given in the form of weighted Dynkin diagrams. A nilpotent element is called \emph{distinguished} iff $\textrm{dim}(\g_0)=\textrm{dim}(\g_2)$ in the Dynkin grading associated to $e$. It follows then that $\g_1=0$ \cite{ELC}.
\begin{examp}\label{principal:nilp}
The principal nilpotent orbit in $\g$ is the  unique nilpotent orbit of codimension $r$(=rank $\g$). Any representative $e$ of this nilpotent orbit is regular, i.e. the centralizer of $e$ in $\g$ is abelian and of dimension $r$. The Dynkin grading is the only good grading associated to $e$. The principal nilpotent orbit is a distinguished nilpotent orbit.
\end{examp}

Throughout the paper we assume all gradings are good with a fixed good element
denoted by $e$.

\subsection{Bi-Hamiltonian reduction for a nilpotent element }
\label{natural bih} Take on $\lop\g$ the Poisson pencil
\eqref{Lie:poisson} with $a\in \cc$ a homogenous element of minimal
degree. Let $\Xi$ be the subset of the set of Casimirs of $P_1$
corresponding to $\lop\nneg~\subset\textrm{Ker } P_1 $. Since $\nneg$ is a Lie subalgebra, it is easy to verify that $\Xi$ is closed under $P_2$. Following Drinfeld and Sokolov \cite{DS} we take as a level surface the affine space
\begin{equation}\s:=\lop \bneg+e
.\end{equation}
The following proposition gives a nice Lie algebra theoretic meaning to the distribution $E$ on $\s$ which is defined by
\begin{equation}E:= P_2(\lop{\nneg})\cap \lop{\bneg}.\end{equation}
\begin{prop}
\begin{equation} E=P_2(\lop\gneg).\end{equation}
\end{prop}
\begin{proof} From \begin{equation}\label{dist:E} E=P_2(\{v\in\lop\nneg:
v_x+[q,v]+[e,v]\in\lop\bneg\textrm{ for } q\in\lop\bneg\})
\end{equation}
and the gradation \eqref{grad1}, it is obvious that $v\in E$ if
and only if $[e,v]\in \lop\bneg$. Since $v\in\lop\nneg$ and ad$~e$
is injective we have $v\in\lop\gneg$.\end{proof}
Fix a transversal space $\cc$ and define the submanifold \begin{equation}Q:=e+\lop\cc\end{equation} of $\s$.
\begin{lem}\label{qistranversal}
The manifold $Q$ is transversal to $E$ on $\s$.
\end{lem}
\begin{proof} We must prove that at any point $q\in \lop\cc$ and
$\dot{s}\in\lop\bneg$ there are $v\in \lop\gneg$ and $\dot{w} \in
\lop\cc$ such that \begin{equation}
\dot{s}=P_2(v)+\dot{w}.\end{equation} We write this equation using
the gradation \eqref{grad1} of $\g$. We obtain
\begin{equation}
\dot{s_i}=v_i'+[e,v_{i-2}]+\dot{w_i}+\sum_k [q_k,v_{i-k}].
\end{equation}
Then for $i=0$ we have
\begin{equation} \dot{s_0}=[e,v_{-2}] +\dot{w_0}\end{equation} which can be solved
uniquely since
\begin{equation}\lop\cc\oplus[e,\lop\gneg]=\lop\bneg.\end{equation} Inductively
in this way for $i<0$ we obtain a recursive relation to determine
$v$ and $\dot{s}$ uniquely.
\end{proof}

Let us explain the procedure of finding the reduced Poisson pencil
following \cite{CP}. We first choose a basis
$\xi_1,...,\xi_n$ for $\g$ with $\xi_1,...,\xi_m$ a basis for
$\cc$ for $m <n$. Let $\xi_1^*,...,\xi_n^* \in \g$ be a dual basis
satisfying $\langle \xi_i|\xi_j^*\rangle=\delta_{ij}$. Then a point in the space $Q$
will have the form $q= q^i \xi_i+ e$. For a covector
$w=(w_1,...,w_m) \in T_q^*Q$ a lift $v \in T_q^*\lop\g$ satisfies
the first condition in lemma  \ref{Self:Consistency} if and only if
\begin{equation}\langle\xi_i|v\rangle=w_i, \qquad i=1,...,m.\end{equation}
From lemma \ref{bi-Hamiltonian reduction} the second
condition gives the constraint \begin{equation} P_\lambda(v) \in
\lop\cc.\end{equation} Using the grading we can prove this lift
is unique. Then the Poisson pencil $P_\lambda^Q$ is given by
\begin{equation}
\dot{q}^i:=\langle P_\lambda (v)|\xi_i^*\rangle.
\end{equation}
Its independence from the choice of a basis follows from lemma
\ref{Self:Consistency}.
\begin{examp}(\textbf{Fractional KdV})\label{FKDVex} Consider $\g=sl_3$ with its standard
representation. We denote by $e_{i,j}$ the fundamental matrix defined by
$(e_{i,j})_{s,t}=\delta_{i,s} \delta_{j,t}$. Take the minimal nilpotent element $e:=e_{1,3}$. It is a good element for
the grading (non Dynkin grading) defined by
\begin{equation}
\widetilde{h}:=\frac{4}{3}e_{1,1}-\frac{2}{3}e_{2,2}-\frac{2}{3}
e_{3,3}
\end{equation}
Take the Poisson tensors
\eqref{Lie:poisson} with $a= e_{2,1}$. Here $\nneg$ is generated
by  $\{ e_{2,1},e_{3,1}\}$ and a point $b\in \s$
will have the form
\begin{equation} b=\left( \begin{array}{ccc}
  * & 0 & 1 \\\,
  * & * & * \\\,
  * & * & *
\end{array}\right).
\end{equation}
We define a transversal space $\cc$ such that a point $q\in Q$ takes the form
\begin{equation}\label{tFKDV}
q=\left( \begin{array}{ccc}
  (\alpha-\beta)q_1 & 0 & 1 \\
  q_2 & - \alpha q_1& 0 \\
  q_4 & q_3 & \beta q_1
\end{array}\right)
\end{equation}
for arbitrary $q_1,...,q_4$ and nonzero constants $\alpha,\beta$.
Then the reduced Poisson pencil $P_\lambda^Q$ has the following form
  \begin{eqnarray*}
    \{q_1(x),q_1(y)\}_\lambda &=& \frac{2\,\delta '(x - y)}{3\,{\alpha }^2} \\
     \{q_1(x),q_2(y)\}_\lambda &=& - \frac{\left( \lambda  + {q_2}(x) \right)\delta (x - y) }{\alpha }  \\
     \{q_1(x),q_3(y)\}_\lambda&=& \frac{\,{q_3}(x)\delta (x - y)}{\alpha } \\
     \{q_1(x),q_4(y)\}_\lambda &=& -\frac{ {\left( \alpha  - 2\,\beta  \right) }^2\,
     {q_1}' \delta (x - y)\, }{3\,{\alpha }^2} - \frac{ {\left( \alpha  - 2\,\beta  \right) }^2\,{q_1}(x)\,\delta '(x - y)  }{3\,{\alpha }^2}
\\&~&- \frac{ \left( \alpha  - 2\,\beta  \right) \,\delta ''(x - y) }{3\,{\alpha }^2}
\\
 \{q_2(x),q_3(y)\}_\lambda &=&\left( \left( 2\,{\alpha }^2 + \alpha \,\beta  - {\beta }^2 \right) \,{{q_1}^2(x)} - {q_4}(x) -
    \left( \alpha  + \beta  \right) \,{q_1}' \right)\delta (x - y)\,
    \\&~& -3\,\alpha \,{q_1}(x)\,\delta '(x - y)+\delta ''(x - y)\\
\{q_2(x),q_4(y)\}_\lambda &=& - \frac{\,\left( 2\,\left( {\alpha
}^2 - \alpha \,\beta  + {\beta }^2 \right)
\,{q_1}(x)\,\left( \lambda  + {q_2}(x) \right)  - \alpha \,{q_2}' \right)\delta (x - y) }{\alpha } \\
         &~&+\frac{\left( \alpha  + \beta  \right) \,\left( \lambda  + {q_2}(x) \right) \,\delta '(x - y)}{\alpha }\\
\{q_3(x),q_4(y)\}_\lambda &=& \frac{\,\left( 2\,\left( {\alpha }^2
- \alpha \,\beta  + {\beta }^2 \right) \,{q_1}(x)\,{q_3}(x) +
      \alpha \,{q_3}' \right) \delta (x - y)}{\alpha }
\\ &~& \frac{\left( 2\,\alpha  - \beta  \right) \,{q_3}(x)\,\delta '(x - y)}{\alpha }
  \end{eqnarray*}
\begin{eqnarray*}
\{q_4(x),q_4(y)\}_\lambda &=&\frac{(2\,{\left( \alpha  - 2\,\beta  \right) }^2\,\left( {\alpha }^2 - \alpha \,\beta  + {\beta }^2 \right)
    \,{q_1}(x)\,{q_1}')\delta (x - y)}{3\alpha^2}\\
    &~& +\frac{(3\,{\alpha }^2\,{q_4}'- 2\,\left( {\alpha }^3 - 3\,{\alpha }^2\,\beta  + 3\,\alpha \,{\beta }^2 - 2\,{\beta }^3 \right) \,
   {q_1}'')\delta (x - y)}{3 \alpha^2}\\
    &~&+2\frac{\,\,\left( 3\,{\alpha }^2\,{q_4}(x) -
      2\,\left( {\alpha }^3 - 3\,{\alpha }^2\,\beta  + 3\,\alpha \,{\beta }^2 - 2\,{\beta }^3 \right) \,{q_1}' \right)\delta '(x - y) }
    {3\,{\alpha }^2}\\
    &~&+2\frac{\,{\left( \alpha  - 2\,\beta  \right) }^2\,\left( {\alpha }^2 - \alpha \,\beta  + {\beta }^2 \right) \,{{q_1}^2(x)}\,
    \delta '(x - y)}{3\,{\alpha }^2}\\&~& -2\frac{\,\left( {\alpha }^2 - \alpha \,\beta  + {\beta }^2 \right)
     \,\delta ^{(3)}(x - y)}{3\,{\alpha }^2}.
\end{eqnarray*}
The vector field defined by a covector $w\in T_q^*Q$ is written in
the form
\begin{equation}\label{FKDV1}
\dot{q}_\lambda=[v,L]
\end{equation}
where $v$ is an extension of $w$  and  $L$ is the matrix operator
\begin{equation}\label{FKDV2}
L =\partial_x+ \left( \begin{array}{ccc}
  (\alpha-\beta)q_1 & 0 & 0 \\
  q_2 & - \alpha q_1& 0 \\
  q_4 & q_3 & \beta q_1
\end{array}\right)+\left(\begin{array}{ccc}
  0& 0 & 1 \\
  \lambda& 0& 0 \\
  0 & 0 & 0
\end{array}\right)
\end{equation}
In the case
$\alpha=2 \beta$, $q_4(x)$ is a Virasoro density, i.e.
\begin{equation}
\{q_4(x),q_4(y)\}_2 =  2\,{q_4}(x)\,\delta '(x - y) + \delta (x -
y)\,{q_4}' - \frac{\delta ^{(3)}(x - y)}{2}
\end{equation}
and the second Poisson bracket is the $W_3^2$-algebra (see e.g. \cite{gDSh1}).
\end{examp}
\begin{rem} Perform the bi-Hamiltonian reduction on $sl_3$ by
 taking the symplectic leaf of $P_1$ defined by setting $a=e_{2,1}+e_{2,3}$
in \eqref{Lie:poisson} and fixing the transversal manifold to have the
form \eqref{tFKDV}. The reduced second Poisson tensor on this
manifold is equal to the one of the example above, (see e.g. \cite{CFMP}).
The form of the operator $L$ will change to
\begin{equation}
L:=\partial_x+\left( \begin{array}{ccc}
  (\alpha-\beta)q_1 & 0 & 0 \\
  q_2 & - \alpha q_1& 0 \\
  q_4 & q_3 & \beta q_1
\end{array}\right)+\left(\begin{array}{ccc}
  0& 0 & 1 \\
  \lambda& 0& 0 \\
  0 & \lambda & 0
\end{array}\right)
\end{equation}
which is the Lax operator considered in \cite{gDSh2} to obtain
integrable hierarchy associated to $W_3^2$-algebra.
\end{rem}
\subsection{Classical W-Algebras from bi-Hamiltonian reduction}

 Classical $W$-algebras and their primary fields  will be obtained for the principal nilpotent element with an appropriate choice of a basis and a transversal subspace constructed using representation theory of $sl_2$-algebra.

Let $e$  be a principal nilpotent element
and $\{h,e,f\}$ is the associated $sl_2$-triple. We denote by $A\subset\g$ the subalgebra generated by this triple. Then we have a decomposition of $\g$ as irreducible $A$-submodules:
\begin{equation}\label{decompForWalg}
\g=A\oplus\oplus_{\alpha=1}^{m} V_{\alpha}.
\end{equation}
Let $n_\alpha +1$ be the dimension of $V_\alpha$. Fix a basis in
$V_\alpha$
\begin{equation} X_{\alpha}^j, \qquad j=0,...,n_\alpha,~ \alpha=1,...,m.\end{equation}  From the representation theory  of $sl_2$ these vectors satisfy the following commutation relations
\begin{eqnarray}\label{gBasisIrSl}
  [h,X_\alpha^j ]&=& (n_\alpha-2j) X_\alpha^j \\
 \nonumber [f,X_\alpha^j] &=& (j+1) X_\alpha^{j+1} \\
  \nonumber [e,X_\alpha^j] &=& (n_\alpha-j+1)X_\alpha^{j-1}.\end{eqnarray}
It is easy to prove that $\cc=~\textrm{Ker}(\ad~f)$ is a transversal subspace associated to $e$. We will apply the generalized bi-Hamiltonian reduction using this transversal subspace.
Let
\begin{equation}
v:=v_j^\alpha X_\alpha^j+ v_h h+v_e
e+v_f f
\end{equation}
be a general covector in $\lop\g ^*$ and
\begin{equation}
q:=q^\alpha X_\alpha^{n_\alpha}+ q_f f+ e
\end{equation}
a point in $Q$. Using \eqref{gBasisIrSl} the second Poisson tensor at $q\in Q$  reads
\begin{eqnarray}\label{hard:eq}
 \nonumber  P_2(v) &=& [\frac{d}{dx}+q^\alpha X_\alpha^{n_\alpha}+ q_f f+e,v_j^\alpha X_\alpha^j+ v_h h+v_e
e+v_f~ f]  \\
  &=& \Psi + (v_j^\alpha)_x ~X_\alpha^j+  (v_h)_x
  ~h+(v_e)_x~
e+(v_f)_x~ f + \\
  \nonumber &~& n_\alpha ~q^\alpha~ v_h~ X_\alpha^{n_\alpha}-q^\alpha~ v_e ~X_\alpha^{n_\alpha-1}+
    (j+1)~ q_f~ v_{j}^{\alpha}~ X_\alpha^{j+1}+ 2 v_h~q_f~ f \\
  \nonumber &~& - q_f~ v_e~  h+
  (n_\alpha-j+1)~ v_j^\alpha~ X_\alpha^{j-1}-
  2 v_h~ e+ v_f~h
\end{eqnarray}
where \begin{equation} \Psi=q^\alpha v_j^\lambda
[X_\alpha^{n_\alpha},X_\lambda^j].
\end{equation}
To find the reduced Poisson tensor $P_2^Q$ one
must solve the recursion relations equating to zero the
coefficients of $X_\alpha^{j}\textrm{ , for }
j=0,...,n_\alpha-1 $ and of $e$ and $h$ (using the procedure explained after lemma \ref{qistranversal}).
\begin{prop}
 The brackets with $q_f$ will be given as follows
\begin{eqnarray}
\quad\{q_f(x),q_f(y)\}&=& - c_1\big
(\frac{1}{2} \delta^{'''}(x-y)+ 2 q_f(x)
\delta^{'}(x-y)\\\nonumber &~&+
(q_f)_x \delta(x-y)\big )\\
\quad\{q^\alpha(x),q_f(y)\}&=& c^\alpha\big(
q_x^\alpha\delta(x-y)+ \frac{(n_\alpha+2)}{2}~ q^\alpha(x)
\delta^{'}(x-y)\big),
\end{eqnarray}
where $c_1$ and $c^\alpha$ are some constants depending on  the
choice of the basis (they are unique up to multiplication of $X_\alpha^0$ by a nonzero constant).
\end{prop}
\begin{proof} The main idea of the proof is to study the
contribution of $v_e$ and its derivatives on the solutions of \eqref{hard:eq}. This  gives the Poisson brackets with $q_f(x)$.
First we put $v_0^\alpha=0,~\alpha=1,\ldots,m$. It follows easily from Dynkin grading
that $$v_i^\alpha=0, ~~i=1,...,{n_\alpha\over 2}, ~\forall \alpha$$(recall $n_\alpha ~\forall \alpha$ is even for principal nilpotent elements).
 It follows that the expansion of $\Psi$ does not contain $h, f$ or $e$.  Therefore equating the coefficient of $e$ to zero we have
\begin{equation}
v_h=1/2 (v_e)_x
\end{equation}
and the coefficient of $h$  gives
\begin{equation}
 v_f = v_e q_f-1/2
(v_e)_{xx}
\end{equation}
and the coefficient of $f $
reads
\begin{eqnarray}
({v}_f)_x&+&2 {v}_h {q}_f =\\\nonumber -
1/2({v}_e)_{xxx}&+& 2({v}_e)_x
q_f+  {v}_e ({q}_f)_x
\end{eqnarray}
Observe that $v_e$ appears explicitly only when equating the coefficient of $X_\alpha^{n_\alpha-1}$ to zero which gives
\begin{equation}\label{lastterm}
v_{n_\alpha}^{\alpha}= q^\alpha v_e+ \textrm{other terms}
\end{equation}
 The next step is using the fact that ker$(\ad~f)$ is abelian subalgebra (since  $f$ is a principal nilpotent element) to rewrite
\begin{equation} \Psi=q^\alpha v_j^\lambda
[X_\alpha^{n_\alpha},X_\lambda^j], \quad j\neq n_\lambda.
\end{equation}
Thus solving the equation \eqref{hard:eq} recursively  we have $$v_j^\alpha=0,~~j={n_\alpha\over 2}+1,\ldots, n_\alpha$$
Then we have
\begin{equation}\label{lastterm}
v_{n_\alpha}^{\alpha}= q^\alpha v_e
\end{equation}
Finally the coefficient of $X_\alpha^{n_\alpha}$  leads to the expression
\begin{equation}({v}_{n_\alpha}^\alpha)_x+
n_\alpha q^\alpha {v}_h
 = q_x^\alpha {v}_e + \frac{(n_\alpha+2)}{2}
q^\alpha ({v}_e)_x
\end{equation}
We substitute $\delta^k(x-y)$ for $\partial_x^k(v_e)$, and the proof is complete.
\end{proof}
 %%%%%%%%%%%%%%%%%%%%%%%%%%%%%%
 Thus we proved the reduced second Poisson brackets to be the classical $W$-algebra as defined in \cite{BalFeh} where
${q}_f(x)$ is a Virasoro density and $q^\alpha(x)$ are
primary fields of weights $\frac{c^\alpha}{2}{(n_\alpha+2)}$.

In \cite{BalFeh} they obtained the same brackets from the Drinfeld-Sokolov reduction associated to $e$ and the transversal space $\cc$. We will discuss the Drinfeld-Sokolov reduction and its relation with bi-Hamiltonian reduction in the next section.
\begin{rem}
In a similar manner one can obtain the Virasoro density using arbitrary distinguished nilpotent element $e'$ (see also [2]). Our methods fail to produce the primary fields since the Poisson bracket with Virasoro density will depend on the structure constants of the transversal space $\cc'=$ ker$(\ad~f')$ where $\{e',h',f'\}$ is the associated $sl_2$-triple.
\end{rem}

\section{Drinfeld-Sokolov reduction}

In this section we will recall briefly the Drinfeld-Sokolov reduction which is another procedure to obtain a bi-Hamiltonian manifold. In the next section we will show its equivalence  to our bi-Hamiltonian reduction. We use the notations and terminology of section \ref{notation}.

Let us  denote by $\s$ the manifold consisting of operators of the form
\begin{equation}\label{op:DS} L=\frac{d}{dx}+b+e\qquad\textrm{where
} b\in \lop\bneg .\end{equation} The adjoint group $\gauge$ of
$\lop\gneg$ acts on $\s$ by
\begin{equation}
 (n,L)\to \e n L \textrm{ for all }
n\in \lop\gneg \textrm{ and } L\in \s. \end{equation}
\begin{prop}
For any operator $L\in \s$ there is a unique element $s \in
\lop\gneg$ such that the operator $\lcan=\exp {\ad {~s}} L$ has the form
\begin{equation}\label{op:fixing} \lcan:=\frac{d}{dx} +q+ e
\end{equation}
where $q \in \lop\cc$. The entries of $q$ are generators of the ring
$R$ of differential polynomials invariant under the action of the
group $\gauge$ on $\s$.
\end{prop}
\begin{proof} We write
$\lcan=\e{s} L$ in the grading associated to $e$. Then inductively  for
$i\leq 0$ the equation has the form
\[ b_i+[e,s_{i-2}]=...\] where the right-hand side is a
differential expression in $q$ and $s$ of the degree greater than
$i$. The result follows by noticing that \[ \cc_i\oplus \ad e
(\gneg_{i-2})=\bneg_i.\]
\end{proof}
From the above lemma  we define the space
$\widetilde{Q}:=\s/\gauge$. The set $\mathcal{R} $ of functionals
on $\widetilde{Q}$ can be realized as functionals on $\s$
which have densities in the ring $R$. Consider the space
$\s$ as a subspace of $\lop \g$. Then for a functional $H$ on $\s$ we
define the gradient $\delta H (q)$ to be the unique element in
$\lop\bpos$ such that
\begin{equation}
\frac{d}{d\theta}H(q+\theta
\dot{s})\mid_{\theta=0}=\int_{s^1}\langle\delta H|\dot{s}\rangle
~~~\textrm{for all } \dot{s}\in \lop\bneg
\end{equation}
and \begin{equation} \int_{s^1}\langle\delta H|\dot{s}\rangle=0
~~~\textrm{for all } \dot{s}\in \lop\nneg.\end{equation}
 We define on $\lop\g$ the
Poisson pencil \eqref{Lie:poisson} with $a\in\g$ a homogenous element of minimal
degree. Then this Poisson pencil $P_\lambda$ is reduced on
$\widetilde{Q}$ using the following
\begin{lem}
$\mathcal{R}$ is a closed subalgebra with respect to the Poisson
pencil $P_\lambda$.
\end{lem}
\begin{proof} Note that if \begin{equation}L=\frac{d}{dx}+ q+ e\in \s \end{equation}and
 \begin{equation}\widetilde{L}:=\frac{d}{dx}+\widetilde{q}+e=\e{n} L\end{equation}
 then for
 $F\in \mathcal{R}$ we have $\delta F(\widetilde{q})= \e{(-n)} \delta
 F(q)$. The proof is easily obtained by using any faithful matrix representation of $\g$.
The result follows by substituting into the bracket
 \begin{equation}
\{F,H\}_\lambda(q)=\int_{s^1}\langle[\delta H,\delta
F]|\frac{d}{dx}+ q+\lambda a \rangle
 \end{equation}
 and using the invariance of the bilinear form $\langle.|.\rangle$
 under the adjoint action.
\end{proof}
\subsection{Drinfeld-Sokolov and bi-Hamiltonian reductions}

\label{NDS}
In this section we will be mainly following the spirit of \cite{Pedroni2}.

\begin{thm}(Marsden-Weinstein reduction)\label{mar:wein}
Let $M'$ be a Poisson manifold with Poisson tensor $P'$, let
$G'$ be a Lie group and $\g'$ its Lie algebra; suppose that $G'$
acts on $M'$ by a Hamiltonian action $\Psi$, with momentum map
$J:M'\rightarrow {\g'}^*$, i.e. for every $\xi\in \g'$ the
fundamental vector field $X_\xi$ is a Hamiltonian vector field
,
\begin{equation} X_\xi=P' \textrm{d} H_\xi\end{equation} with Hamiltonian
$H_\xi(m)=(J(m),\xi)$. Suppose the momentum map $J$ to be
$\Ad^*$-equivariant, i.e. \begin{equation}
J(\Psi_g(u))=\Ad_g^* J(u) ~ \hbox{for all }g \in G'.\end{equation} Let
$\mu\in \g'$ be a regular value of $J$, so that  $\s'
=J^{-1}(\mu)$ is a submanifold of $M'$, and let $D'$ be the
tangent distribution to the orbits of $\Psi$. Then  the triple
$(M',\s',D')$ is Poisson-reduced using the Marsden-Ratiu
reduction theorem \ref{mar:rat}. The quotient manifold turns
out to be $N'=J^{-1}(\mu)/ {G'}_\mu$, where ${G'}_\mu$ is the
isotropy group of $\mu$.
\end{thm}

Let $M$ be the space of operators of the
form
\begin{equation}\label{op:gen} Z=\frac{d}{dx}+q\qquad\textrm{where
} q\in \lop\g.
\end{equation}

The adjoint group $\ngauge$ of $\lop\nneg$ acts on $M$ by
\begin{equation}
 (n,Z)\to \e n Z \textrm{ for all }
n\in \nneg \textrm{and } Z\in M .\end{equation}
 Introduce on $M$ the bi-Hamiltonian structure \eqref{Lie:poisson} with $a \in \cc$ a homogenous element
 of minimal degree.
\begin{prop}
The action of $\ngauge$ on $M$ with Poisson tensor $P_\lambda$ is
Hamiltonian for all $\lambda$. It admits a momentum map $J$ to be
the projection
\[J:\g\to\gpos.\] Moreover, $J$ is $\Ad^*$-equivariant.
\end{prop}
\begin{proof} We consider a faithful matrix representation of $\g$. Then the
action on $M$ has the form\begin{equation} \Psi_n:q\to n q n^{-1}-
n_x n^{-1}\qquad q\in \lop\g , n\in \ngauge.\end{equation} For
$\xi\in \lop\nneg$ the fundamental vector field will have the form
\begin{eqnarray}\label{action distri}
X_{-\xi} &=&\frac{d}{dt}\big(\textrm{exp}(-t\xi)~ q
~\textrm{exp}(t\xi)-(\textrm{exp}(-t\xi))_x\textrm{exp}(t\xi)\big)\\
\nonumber &=& \xi_x+ [q,\xi].
\end{eqnarray}
Define the functional \begin{equation}H_\xi(q)=\int \langle
q,\xi\rangle=\int \langle J(q),\xi\rangle.\end{equation}
 Then\begin{equation}
 P_\lambda
\delta H_\xi=\xi_x + [\xi,q+\lambda
a]=\xi_x+[\xi,q],\end{equation} which proves the action is
Hamiltonian. The momentum map is $\Ad^*$-equivariant iff
\[J(\Psi_n(q))=\Ad_n^* J(q).\] Since the moment map is just the
projection we have
\begin{eqnarray*}
J(\Psi_n(q))&=& J(n q n^{-1}-n_x n^{-1})\\&=& J(n q n^{-1})\\
&=& J(n J(q) n^{-1}) \\
&=& \Ad_n^* J(q)
\end{eqnarray*}
where the last equality follows from the fact that the coadjoint
action of $\ngauge$ on  $\gpos \simeq {\nneg}^*$ is given by
\begin{equation}\Ad_n^* v= J(n v n^{-1}),\qquad v\in
{\nneg}^*.\end{equation}
\end{proof}
We take the nilpotent element $e$ as a regular value of $J$. Define the space \begin{equation}\s:=J^{-1}(e)=\frac{d}{dx}+\lop{\bneg}+e\end{equation} and let
 $D$ denote the distribution defined by the group action
\begin{equation}D:=P_\lambda(\lop\nneg)=P_2(\lop\nneg).\end{equation}
Let $E:=D\cap TS$. Then
$P_\lambda,\s$ and $D$ satisfy Marsden-Ratiu theorem \cite{MR}.

According to theorem \ref{mar:wein}, $P_\lambda$ is reduced on the space
\begin{equation}\widetilde{Q}:=\s/\gauge\end{equation} where
$\gauge\subset\ngauge$ is the isotropy subgroup of $e$ under the
action of $\ngauge$. From the properties of the grading $\gauge$ is
the adjoint group of $\gneg$. This obviously leads to  Drinfeld-Sokolov
reduction.

Now we use corollary \ref{mar:rat:cor}. Define the  space
\begin{equation}Q:=\frac{d}{dx}+\lop\cc+e.\end{equation} Then $Q$ is
transversal to the distribution $E:=D\cap TS=P_2(\lop\bneg)$ on
$S$ by lemma \ref{qistranversal} and \ref{dist:E}. This gives the generalized bi-Hamiltonian reduction.

Thus we have proved the following
\begin{thm}
The generalized Drinfeld-Sokolov  and generalized bi-Hamiltonian reductions are equivalent in the sense that they satisfy the Marden-Ratiu theorem with the same Poisson pencil $P_\lambda$, the submanifold $\s$ and the distribution $D$.
\end{thm}

As we mentioned in the introduction, in the special case of principal nilpotent element the equivalence is obtained in \cite{Pedroni2}. Using generalized bi-Hamiltonian reduction the proof is more simpler even in this case.

\section{Applications to Frobenius manifolds}

Let $M$ be a manifold with local coordinates $(U^1,...,U^n)$. On the loop space $\lop M$ a local Poisson bracket can be written in the form \cite{DZ}
\begin{equation} \label{genLocPoissBra}\{U^i(x),U^j(y)\}=
\sum_{k=-1}^\infty \epsilon^k \{U^i(x),U^j(y)\}^{[k]}.
 \end{equation}
 Here $\epsilon$ is just a parameter and
 \begin{equation}\label{genLocBraGen}
\{U^i(x),U^j(y)\}^{[k]}=\sum_{s=0}^{k+1} A_{k,s}^{i,j}
\delta^{(k-s+1)}(x-y),
 \end{equation}
where  $A_{k,s}^{i,j}$ are homogenous polynomials in $\partial_x^j
U^i(x)$ of degree $s$ (we assign
$\partial_x^j U^i(x)$ degree $j$). The first terms can be written
as follows
\begin{eqnarray}
  \{U^i(x),U^j(y)\}^{[-1]} &=& F^{ij}(U)\delta(x-y) \\
  \{U^i(x),U^j(y)\}^{[0]} &=& g^{ij}(U) \delta' (x-y)+ \Gamma_k^{ij}(U) U_x^k \delta (x-y)
\end{eqnarray}
where $F^{ij}$, $g^{ij}$ and $\Gamma_k^{ij}$ are smooth functions
in $U^i$. The matrix $F^{ij}$ defines a Poisson structure on
$M$. If $F^{ij}=0$ then  $\{U^i(x),U^j(y)\}^{[0]}$ defines a
Poisson bracket on $\lop M$ known as Poisson bracket of
hydrodynamic type. By nondegenerate Poisson bracket of
hydrodynamic type we mean the metric $g^{ij}$ is nondegenerate. In
this case its inverse defines a flat metric on the tangent space $TM$ and
$\Gamma_k^{ij}$ are the contravariant Levi-Civita coefficients of
$g^{ij}$ \cite{DN}. Assume there is a bi-Hamiltonian structure on $\lop M$ defined by Poisson tensors $P_1$ and $P_2$. Suppose $P_1$ and $P_2$ admit a nondegenerate Poisson brackets of hydrodynamics type and $\det(g_2^{ij}-\lambda g_1^{ij})\neq 0 ,~\hbox{for generic } \lambda$. Then by definition \cite{DFP} $g_{1;2}^{ij}$ form a
 flat pencil of metrics. Under some
assumption of quasihomogeneity and regularity a flat pencil of
metrics is equivalent to a Frobenius structure
on $M$ \cite{DFP}. In the notations of \eqref{frob} from a Frobenius structure on $M$ the flat pencil of metric is
found from the relations \begin{eqnarray} \eta^{ij}&=&g_1^{ij} \\
\nonumber g_2^{ij}&=&(d-1+d_i+d_j)\eta^{i\alpha}\eta^{j\beta}
\partial_\alpha
\partial_\beta \mathbb{F}
\end{eqnarray}

\subsection{Dirac reduction}

In this section we write the formulas for a Dirac reduction  of a Poisson bracket on the loop space $\lop M$ to a loop space $\lop N$ of a suitable submanifold $N\subset M$. We use the notations introduced in the beginning of this section.

It is well  known that the Dirac reduction of Poisson bracket of hydrodynamic types  may result  a nonlocal Poisson
bracket \cite{FerNLoc}. We obtain a Dirac reduction for a local Poisson bracket which does not admit a dispersionless limit. The resulting Poisson bracket is local.

Let $N$ be a submanifold of $M$ of dimension $m$.  Assume $N$ is defined by the equations $U^\alpha=0$ for $\alpha=m+1,...,n$. We introduce three types of indexes; capital letters $I,J,K,...=1,..,n$,
small letters $i,j,k,...=1,....,m$ which parameterize the
submanifold $N$ and Greek letters
$\alpha,\beta,\delta,...=m+1,...,n$.
\begin{prop} Assume  the matrix
$F^{\alpha \beta}$ is nondegenerate. Then Dirac reduction is well defined on $\lop {N}$ and gives a local Poisson bracket.
\end{prop}
\begin{proof}
 Let $\f$ be a Hamiltonian functional on $\lop {M}$. Then the Hamiltonian flows have the equation
 \begin{equation}\label{DRed1}
 U_t^I=B^{IJ} \frac{\delta \f}{\delta U^J}
 \end{equation}
 where \[ B^{IJ}=\epsilon^k A^{IJ}_{k,s} {d^{k-s+1} \over dx^{k-s+1}}.\]

 Then Dirac equation on $N$ will have the form
 \begin{eqnarray}\label{Dred2}
 U_t^i&=& B^{iJ} \frac{\delta \f}{\delta U^J}+ \int \{U^i,U^\beta\}
 C_\beta(y) dy \\
\nonumber &=& B^{ij} \frac{\delta \f}{\delta U^j}+
B^{i\beta}(\frac{\delta \f}{\delta U^\beta}+C_\beta)
\end{eqnarray}
where $C_\beta(y)$ can be found from the equation
\begin{eqnarray}\label{Dred3}
 0=U_t^\alpha &=& B^{\alpha J} \frac{\delta \f}{\delta U^J}+ \int \{U^\beta,U^\beta\}
 C_\beta(y) dy \\
\nonumber &=& B^{\alpha j} \frac{\delta \f}{\delta U^j}+
B^{\alpha\beta}(\frac{\delta \f}{\delta U^\beta}+C_\beta).
\end{eqnarray}
In powers of $\epsilon$ this equation reads
 \begin{multline}\label{Dred4}
     -(\epsilon^{-1} B_{-1}^{\alpha i}+B_{0}^{\alpha i}+\epsilon B_{1}^{\alpha i}+...)\frac{\delta \f}{\delta U^i}=\\
     (\epsilon^{-1} B_{-1}^{\alpha \beta}+B_{0}^{\alpha \beta}+\epsilon B_{1}^{\alpha \beta}+...)(\epsilon^{-1} C_\beta^{-1}+ (\frac{\delta \f}{\delta
        U^\beta}+C_\beta^0)+\epsilon C_\beta^1+...).
\end{multline}
We will solve this equation  recursively. We depend on the fact that the matrix $B_{-1}^{\alpha \beta}=F^{\alpha \beta}$ is  invertible. Then the  coefficients of $\epsilon^{-2}$
gives
\begin{equation}\label{Dred5}
0= F^{\alpha\beta} C_\beta^{-1} ~ \Rightarrow C_\beta^{-1}=0
\end{equation}
and the coefficient of $\epsilon^{-1}$
\begin{eqnarray}\label{Dred6}
  F^{\alpha i} \frac{\delta \f}{\delta U^i} &=& -F^{\alpha \beta} (\frac{\delta \f}{\delta U^\beta}+C_\beta^0) \\
  \nonumber (\frac{\delta \f}{\delta U^\beta}+C_\beta^0)&=& -
  F_{\beta \alpha} F^{\alpha i} \frac{\delta \f}{\delta U^i}
\end{eqnarray}
where $F_{\beta\alpha}$ is the inverse of $F^{\beta \alpha}$. The
constant term in $\epsilon$ leads to
\begin{eqnarray}\label{Dred7}
-B_0^{\alpha i}\frac{\delta \f}{\delta U^i}&=& B_0^{\alpha
\beta}(\frac{\delta \f}{\delta U^\beta}+C_\beta^0) + F^{\alpha
\beta} C_\beta^1 \\
C_\beta^1&=& F_{\beta \alpha}(-B_0^{\alpha i} \frac{\delta
\f}{\delta U^i}+B_0^{\alpha \varphi} (F_{\varphi \gamma}F^{\gamma
i} \frac{\delta \f}{\delta U^i})). \end{eqnarray}
Note that $B_0$
are differential operators. We  continue in this way to find all the elements $C_\beta^i$
for $i>1$. In fact

\begin{equation}
\begin{split}
            C_\beta^s= & F_{\beta\alpha}\big(-B_{s-1}^{\alpha i} {\delta\f\over \delta U^i}-B_{s-1}^{\alpha\varphi}({\delta\f\over\delta U^{\varphi}}+C_\varphi^0)  \\
              &-B_{s-2}^{\alpha\varphi}C_\varphi^{1}-B_{s-3}^{\alpha\varphi}C_\varphi^2-...\big)
          \end{split}
\end{equation}
Therefore we get a differential operators acting on the vector $\frac{\delta
\f}{\delta U^i}$. Finally, we substitute the values of $C_\beta^i$ in
the equations \eqref{Dred2}. We get a local Poisson bracket on
$\lop {N}$. This ends the proof.\end{proof}

The Hamiltonian equations on $N$ read
\begin{eqnarray}
  \nonumber U_t^i &=& \sum_{k=-1}^\infty \epsilon^{k} B_{k}^{ij} \frac{\delta \f}{\delta U^j}+ \\
  \nonumber &~& \big(\sum_{k=-1}^\infty \epsilon^{k} B_{k}^{i\beta}\big) \big((\frac{\delta \f}{\delta U^\beta}+ C_\beta^0)+ \epsilon C_\beta^1+...\big)\\
  \nonumber &=&\sum_{k=-1}^\infty \epsilon^{k} B_{k}^{ij} \frac{\delta \f}{\delta U^j}+ \\
  \nonumber &~& \big(\sum_{k=-1}^\infty \epsilon^{k} B_{k}^{i\beta}\big) \big( -
  F_{\beta \alpha} F^{\alpha i} \frac{\delta \f}{\delta U^i}+ \epsilon F_{\beta \alpha}(-B_0^{\alpha i} \frac{\delta
\f}{\delta U^i}+B_0^{\alpha \varphi} (F_{\varphi \gamma}F^{\gamma
i} \frac{\delta \f}{\delta U^i}))+ O(\epsilon^2)\big).
\end{eqnarray}
Hence if we write the Poisson bracket on $N$ in the form
\begin{eqnarray}
  \{U^i(x),U^j(y)\}^{[-1]} &=& \widetilde{F}^{ij}\delta(x-y) \\
  \{U^i(x),U^j(y)\}^{[0]} &=& \widetilde{g}^{ij} \delta' (x-y)+ \widetilde{\Gamma}_k^{ij} U_x^k \delta (x-y).
\end{eqnarray}
We have
\begin{equation}
\widetilde{F}^{ij}=(F^{ij}-F^{i\beta} F_{\beta\alpha} F^{\alpha j}
)\end{equation}
and
\begin{equation}
\widetilde{g}^{ij}= g^{ij}-g^{i\beta} F_{\beta\alpha}F^{\alpha
j}+F^{i \beta}F_{\beta \alpha} g^{\alpha \varphi} F_{\varphi
\gamma} F^{\gamma j}-F^{i\beta} F_{\beta \alpha} g^{\alpha j}
\end{equation}
from the coefficient of $\epsilon^{-1}$ and of $\epsilon^0$ respectively.

\begin{rem}
     As expected the formula of
$\widetilde{F}^{ij}$ coincide with Dirac reduction of the finite dimensional
Poisson bracket defined by $F^{IJ}$ on $M$ to $N$.
\end{rem}

\begin{cor}\label{rem:dirac}
Assume $U^i,~i=1,...,n$ are  Casimirs of the Poisson bracket $F^{IJ}$ on $M$. Then $F^{i\alpha}=0$. Hence the reduced Poisson bracket on $\lop N$ reads \begin{eqnarray} \widetilde{F}^{ij}&=& 0\\\nonumber  \widetilde{g}^{ij}&=&g^{ij}.
 \end{eqnarray}

\end{cor}

\subsection{Algebraic Frobenius manifolds from  Lie algebra $F_4$}

Our aim is to obtain a Frobenius manifold from a reduced Lie-Poisson pencil. We apply generalized bi-Hamiltonian
reduction to distinguished nilpotent elements of the Lie algebra $F_4$. There are four
distinguished nilpotent orbits on $F_4$. They correspond to regular conjugacy classes in Weyl group $\w_{F_4}$ \cite{DelFeher}.

 Let $e$ be a distinguished nilpotent element in $F_4$. We apply the bi-Hamiltonian reduction to $e$ with  Dynkin grading. Let $P_\lambda^Q$ denote the reduced Poisson pencil on the loop space $\lop Q$. If $e$ is not principal then the reduced Poisson pencil $P_\lambda^Q$ does not admit a dispersionless limit, i.e. the leading term of $P_\lambda^Q$ of degree $-1$ does not vanish. We apply Dirac reduction to a submanifold $N\subset Q$ such that all the pencil $P_\lambda^Q$ is reduced. The new Poisson pencil $P_\lambda^N$ on $\lop N$ gives a  nondegenerate Poisson pencil of hydrodynamic type. The example below illustrates this procedure in details.

    The distinguished nilpotent elements in $F_4$ give four non isomorphic algebraic Frobenius manifolds of dimension 4. Two of them  give a polynomial Frobenius manifolds isomorphic to Frobenius structure on the orbit spaces of Coxeter
group of type $F_4$ and $B_4$ \cite{DCG}. One
of the remaining distinguished nilpotent orbits is likely to give
algebraic Frobenius structure isomorphic to the one found in
\cite{PAV} by applying  Drinfeld-Sokolov  reduction on Lie
algebra of type $D_4$. We obtain the same result by applying the generalized bi-Hamiltonian reduction to $D_4$. We end with one class of nilpotent
elements which give a new algebraic Frobenius manifold.

\begin{examp}(Algebraic Frobenius manifold)
Denote by $\Psi$ the set of roots of the Lie algebra $F_4$. For the following computations we use
the minimal representation of $F_4$ \cite{CAR}. Assume $X_\alpha $ with $H_\alpha\in \h$, $\alpha\in \Psi$, form
Weyl-Chevalley basis of $F_4$.  We apply the  generalized bi-Hamiltonian reduction with the nilpotent
element
\end{examp}
\begin{eqnarray}
\nonumber e&:=&{X_{{{\alpha }_2}}} + {X_
    {{{\alpha }_1} + {{\alpha }_2}}} +
  {X_{{{\alpha }_2} + {{\alpha }_3}}} +
  {X_{{{\alpha }_1} + {{\alpha }_2} +
      {{\alpha }_3}}} +
  {X_{{{\alpha }_2} + 2\,{{\alpha }_3}}}\\&~& +
  {X_{{{\alpha }_1} + {{\alpha }_2} +
      2\,{{\alpha }_3}}} + {X_{{{\alpha }_4}}} +
  {X_{{{\alpha }_3} + {{\alpha }_4}}}
\end{eqnarray}
which is a representative  of the nilpotent orbit $F_4(a_2)$ in the notations of \cite{COLMC}. We fix the associated Dynkin grading and define the first Poisson bracket with $a={X_{-2\,{{\alpha }_1} - 3\,{{\alpha }_2} - 4\,{{\alpha }_3}
- 2\,{{\alpha }_4}}}$. Define the transversal manifold $Q$ to be of the form
\begin{eqnarray}
\label{tranF4a2}
 \nonumber   q&=&{U_2}\,{X_{-{{\alpha }_2} - 2\,{{\alpha }_3}}} + {U_3}\,{X_{-{{\alpha }_1} - {{\alpha }_2} - 2\,{{\alpha }_3}}} +
  {U_1}\,{X_{-{{\alpha }_1} - {{\alpha }_2} - {{\alpha }_3}}} \\
  \nonumber &~& +  {U_8}\,{X_{-2\,{{\alpha }_1} - 3\,{{\alpha }_2} - 4\,{{\alpha }_3} - 2\,{{\alpha }_4}}} +
  {U_7}\,{X_{-{{\alpha }_1} - 3\,{{\alpha }_2} - 4\,{{\alpha }_3} - 2\,{{\alpha }_4}}}\\
   \nonumber &~& +{U_5}\,{X_{-{{\alpha }_1} - {{\alpha }_2} - 2\,{{\alpha }_3} - 2\,{{\alpha }_4}}} +
  {U_4}\,{X_{-{{\alpha }_1} - {{\alpha }_2} - 2\,{{\alpha }_3} - {{\alpha
  }_4}}}\\
  &~& +
  {U_6}\,{X_{-{{\alpha }_1} - 2\,{{\alpha }_2} - 3\,{{\alpha }_3} - 2\,{{\alpha
  }_4}}}
.\end{eqnarray}

Write the reduced Poisson pencil in the notations of \eqref{genLocBraGen}. Then the coefficient $F_\lambda^{ij}$ of
$\epsilon^{-1}$ does not vanish. Indeed, the coefficient
$F_1^{ij}$ of the first Poisson bracket reads
\begin{equation}F_1^{ij}=\left(\begin{array}{cccccccc}
0&0&0&0&0&\frac{1}{6}&0&0\\ & & & & & & & \\
 *&0&0&0&0&\frac{5}{72}&0&0\\ & & & & & & & \\
  *&*&0&0&0&- \frac{11}{72} &0&0\\ & & & & & & & \\
  *&*&*&0&- \frac{4}{5} &0&\frac{10\,{U_1} + 23\,{U_2} + 5\,{U_3}}{120}&
   \frac{-6\,{U_1} + 43\,{U_2} + {U_3}}{120}\\ & & & & & & & \\
  *&*&*&*&0&\frac{-17\,{U_1} - 198\,{U_2} - 18\,{U_3}}{180}&
   \frac{-{U_4}}{32}&\frac{-{U_4}}{160}\\  & & & & & & & \\
   * &* &*&*&*&0& F_1^{6,7}& F_1^{6,8}\\  & & & & & & & \\
       *&*&*&*&*&*&0&\frac{5\,{U_1}\,{U_4} - 24\,{U_2}\,{U_4}}{1920}\\ & & & & & & & \\
 *&*&*&*&*&*&*&0\end{array}\right)
\end{equation}
where\begin{eqnarray*}
 F_1^{6,7}&=& \frac{-15\,{{U_1}}^2 - 380\,{U_1}\,{U_2} -
700\,{{U_2}}^2 - 60\,{U_1}\,{U_3} - 328\,{U_2}\,{U_3} -
60\,{{U_3}}^2 + 150\,{U_5}}
    {5760}\\
F_1^{6,8}&=&\frac{69\,{{U_1}}^2 + 772\,{U_1}\,{U_2} -
1356\,{{U_2}}^2 - 28\,{U_1}\,{U_3} - 424\,{U_2}\,{U_3} -
76\,{{U_3}}^2 +
      110\,{U_5}}{5760}
.\end{eqnarray*} Our next aim is to use Dirac reduction. For this we introduce the coordinates
\begin{eqnarray*}\label{try}
 \nonumber W_1 &=& \frac{{U_1}}{2} + {U_2} + {U_3} \\
  \nonumber  W_2&=& 11\,{U_2} + 5\,{U_3}\\
   \nonumber W_3 &=& \frac{{U_1}\,\left( -205\,{{U_1}}^2 + 1908\,{U_1}\,{U_2} - 7416\,{{U_2}}^2 + 360\,\left( {U_1} + 2\,{U_2} \right) \,{U_3} +
      360\,{{U_3}}^2 \right) }{1080} \\
   \nonumber   &~& +\frac{5\,{U_1}\,{U_5}}{12} - 2\,{U_2}\,{U_5} + {U_7} - 5\,{U_8}\\
  \nonumber  W_4 &=& \frac{95\,{{U_1}}^3 - 849\,{{U_1}}^2\,{U_2} + 3948\,{U_1}\,{{U_2}}^2 - 195\,{{U_1}}^2\,{U_3} - 120\,{U_1}\,{U_2}\,{U_3} -
    180\,{U_1}\,{{U_3}}^2}{720}\\
   \nonumber  &~& +\frac{{{U_4}}^2}{32} + \left( \frac{-{U_1}}{12} + \frac{19\,{U_2}}{12} + \frac{{U_3}}{12} \right) \,{U_5} + {U_7} + 3\,{U_8} \\
  W_5 &=& U_1 \\
 \nonumber   W_6&=& U_4 \\
 \nonumber   W_7&=& U_5 \\
   \nonumber W_8&=& U_6
\end{eqnarray*}
where the first four are the Casimirs of $F_1^{ij}$. We rewrite the Poisson bracket in the new coordinates. Define the submanifold $N$ given
by
\begin{eqnarray}
\nonumber  W_6&=&W_8=0\\
\nonumber W_5&=& Z\\
\nonumber W_7&=& \frac{-5\,Z^2 + 150\,Z\,{W_1} -
    100\,{{W_1}}^2- 33\,Y\,{W_2} + 64\,{W_1}\,{W_2} -
    7\,{{W_2}}^2}{150}
\end{eqnarray}
where $Z(W_1,...,W_4)$ is a solution of a cubic equation to be given below. It turn out that along $N$ the entries $F_\lambda^{i\alpha}=0,~i=1,...,4,~\alpha=5,...,8$. Hence From corollary \ref{rem:dirac} the Poisson pencil $P_\lambda^Q$ is reduced along $N$. The new Poisson bracket $P_\lambda^N$ gives a flat pencil of metrics. This result a Frobenius structure on $N$. In a flat coordinates $(s_1,s_2,s_3,s_4)$ the  Frobenius structure will have the potential
\begin{eqnarray}\label{Frob1}
\nonumber\mathbb{F}&=& \frac{9\,Z^2\,{{s_1}}^5}{44800} +
\frac{3\,Z^2\,{{s_1}}^4\,{s_2}}{89600} -
  \frac{3\,Z^2\,{{s_1}}^3\,{{s_2}}^2}{89600} -
  \frac{3\,Z^2\,{{s_1}}^2\,{{s_2}}^3}{640000} +
  \frac{153\,Z^2\,{s_1}\,{{s_2}}^4}{89600000}\\\nonumber &~& +
  \frac{1107\,Z^2\,{{s_2}}^5}{4480000000}  +
  \frac{81\,Z^2\,{{s_1}}^2\,{s_3}}{2800} +
  \frac{243\,Z^2\,{s_1}\,{s_2}\,{s_3}}{14000} +
  \frac{729\,Z^2\,{{s_2}}^2\,{s_3}}{280000}\\\nonumber&~& +\frac{409\,Z\,{{s_1}}^6}{2419200}-
  \frac{191\,Z\,{{s_1}}^5\,{s_2}}{1344000}  +
  \frac{187\,Z\,{{s_1}}^4\,{{s_2}}^2}{5376000} +
  \frac{67\,Z\,{{s_1}}^3\,{{s_2}}^3}{13440000}\\ \nonumber &~& -
  \frac{319\,Z\,{{s_1}}^2\,{{s_2}}^4}{179200000} +
  \frac{529\,Z\,{s_1}\,{{s_2}}^5}{13440000000} +
  \frac{1247\,Z\,{{s_2}}^6}{38400000000} +
  \frac{27\,Z\,{{s_1}}^3\,{s_3}}{560}\\ \nonumber &~& -
  \frac{117\,Z\,{{s_1}}^2\,{s_2}\,{s_3}}{5600} +
  \frac{9\,Z\,{s_1}\,{{s_2}}^2\,{s_3}}{56000} +
  \frac{369\,Z\,{{s_2}}^3\,{s_3}}{560000} +
  \frac{243\,Z\,{{s_3}}^2}{70}
  \\
   \nonumber &~& -
  \frac{29459\,{{s_1}}^6\,{s_2}}{580608000}+
  \frac{6089\,{{s_1}}^5\,{{s_2}}^2}{276480000} -
  \frac{254609\,{{s_1}}^4\,{{s_2}}^3}{34836480000} +
  \frac{152263\,{{s_1}}^3\,{{s_2}}^4}{116121600000}\\ \nonumber &~& -
  \frac{300457\,{{s_1}}^2\,{{s_2}}^5}{5806080000000}  -
  \frac{1973651\,{s_1}\,{{s_2}}^6}{174182400000000} +
  \frac{292289\,{{s_2}}^7}{193536000000000} +
  \frac{17\,{{s_1}}^4\,{s_3}}{44800} \\ \nonumber &~&-
  \frac{2647\,{{s_1}}^3\,{s_2}\,{s_3}}{336000} +
  \frac{6059\,{{s_1}}^2\,{{s_2}}^2\,{s_3}}{2240000} -
  \frac{18223\,{s_1}\,{{s_2}}^3\,{s_3}}{33600000} \\ \nonumber &~&+
\frac{11443\,{{s_1}}^7}{174182400}+
  \frac{60131\,{{s_2}}^4\,{s_3}}{1344000000} + \frac{{s_1}\,{{s_3}}^2}{20} +
  \frac{3\,{s_2}\,{{s_3}}^2}{200}\\ &~& - 2\,{s_1}\,{s_3}\,{s_4} +
  \frac{3\,{s_2}\,{s_3}\,{s_4}}{5} + 2\,{s_1}\,{{s_4}}^2
\end{eqnarray}
where $Z$ is a solution of the cubic equation
\begin{eqnarray}
\qquad &~& Z^3 - Z\,\left( \frac{{{s_1}}^2}{48} +
\frac{{s_1}\,{s_2}}{80} +
    \frac{3\,{{s_2}}^2}{1600} \right)\\\nonumber &~& -\frac{{{s_1}}^3}{96} +
\frac{13\,{{s_1}}^2\,{s_2}}{2880} -
  \frac{{s_1}\,{{s_2}}^2}{28800} - \frac{41\,{{s_2}}^3}{288000} -
  \frac{3\,{s_3}}{2}=0
\end{eqnarray}
It is straightforward to check validity of the WDVV equations
for this potential. The quasihomogeneity reads
\begin{equation}
  s_4 \partial_1 F(s)+ s_3 \partial_3 F(s)+\frac{1}{3} s_2\partial_2 F(s)+\frac{1}{3}  s_1\partial_1 F(s) = {7\over 3}
  F(s).
\end{equation}

The four examples of algebraic Frobenius manifolds  obtained from Lie algebra $F_4$ are related to their conjugacy classes in $\w_{F_4}$ as follows. If the regular  conjugacy class is of order $o_4+1$ and the eigenvalues are $\omega^{o_i},~ i=1,...,4$ where $\omega$ is a primitive $(o_4+1)$-th  roots of unity, then the degrees of the corresponding algebraic Frobenius manifold are $o_i+1\over o_4+1$ and the charge is $1-{o_1+1 \over o_4+1}$.

The examples of algebraic  Frobenius manifolds obtained on the Lie algebra $F_4$ suggest that algebraic  Frobenius structures exist for all regular primitive conjugacy classes in Weyl groups.

\bigskip

In a subsequent publication we will consider further examples of
Frobenius structures and integrable hierarchies on bi-Hamiltonian
manifolds produced by applying the reduction methods introduced
in this paper.

 \vskip 0.5truecm \noindent{\bf Acknowledgments.}

The author is extremely grateful to his supervisor B. Dubrovin for
posing him this problem, his constant encouragement and support.
The author also thanks P. Lorenzoni for stimulating discussions and M. Pedroni for his help in writing this paper.
The present work is partially supported by the European Science
Foundation Programme ``Methods of Integrable Systems, Geometry,
Applied Mathematics" (MISGAM), Marie Curie RTN ``European Network
in Geometry, Mathematical Physics and Applications"  (ENIGMA). The
work is also partially supported by Italian Ministry of
Universities and Researches (MUR) research grant PRIN 2006
``Geometric methods in the theory of nonlinear waves and their
applications". The author would like to thank ``The Canon Foundation for
Scientific Research" for supporting his travel to attend ``Poisson 2006" conference.

%%%%%%%%%%%%%%%%%%%%%%%%%%%%%%%%%%%%%
%%%%%%%%%              %%%%%%%%%%%%%%
%%%%%%%%%  References  %%%%%%%%%%%%%%
%%%%%%%%%              %%%%%%%%%%%%%%
%%%%%%%%%%%%%%%%%%%%%%%%%%%%%%%%%%%%%

\end{document}